\documentclass[pra,twocolumn,amsmath,amssymb]{revtex4-1}

\usepackage{epsfig,amsmath}
\usepackage{subfigure}
\usepackage{graphicx}
\usepackage{dcolumn}
\usepackage{stmaryrd}
\usepackage{mathrsfs}
\usepackage{pifont}
\usepackage{amsthm}
\usepackage{amssymb}
\usepackage{bm}
\usepackage{latexsym}
\usepackage[colorlinks=true,linkcolor=blue,citecolor=blue]{hyperref}
\usepackage{color}

\theoremstyle{plain}

\newcommand{\la}{\langle}
\newcommand{\ra}{\rangle}

\newcommand{\ti}{\tilde}
\newcommand{\ga}{\gamma}
\newcommand{\Ga}{\Gamma}

\newcommand{\da}{\dagger}

\newcommand{\al}{\alpha}
\newcommand{\si}{\sigma}

\newcommand{\om}{\omega}

\newcommand{\de}{\delta}

\newcommand{\non}{\nonumber}
\newcommand{\pa}{\partial}

\def\jpb#1{{ J.\ Phys.\ B} {\bf#1}}
\def\jpa#1{{ J.\ Phys.\ A} {\bf#1}}
\def\pra#1{{ Phys.\ Rev. A\/} {\bf#1}}

\def\pre#1{{ Phys.\ Rev. E\/} {\bf#1}}
\def\prl#1{{ Phys.\ Rev.\ Lett.} {\bf#1}}

\def\pla#1{{ Phys.\ Lett. A\/} {\bf#1}}
\def\rmp#1{{ Rev. \ Mod. \ Phys.} {\bf#1}}

\def\njp#1{{ New. J. \ Phys.} {\bf#1}}

\begin{document}

\title{Control relaxation via dephasing: an exact quantum state diffusion study}

\author{Jun Jing$^{1,2}$, Ting Yu$^{3,4}$, Chi-Hang Lam$^{5}$, J. Q. You$^{3}$, and Lian-Ao Wu$^{2}$}\thanks{lianao.wu@ehu.es}

\affiliation{$^{1}$Department of Physics, Zhejiang University, Hangzhou 310027, Zhejiang, China \\ $^{2}$Department of Theoretical Physics and History of Science, University of the Basque Country (EHU/UPV), PO Box 644, 48080 Bilbao, and Ikerbasque, Basque Foundation for Science, 48011 Bilbao, Spain \\ $^{3}$Beijing Computational Science Research Center, Beijing 100193, China \\ $^{4}$Center for Controlled Quantum Systems and Department of Physics and Engineering Physics, Stevens Institute of Technology, Hoboken, New Jersey 07030, USA \\ $^{5}$Department of Applied Physics, Hong Kong Polytechnic University, Hung Hom, Hong Kong, China}

\date{\today}

\begin{abstract}
Dynamical decoupling as one of quantum control strategies aims at suppressing quantum decoherence adopting the popular philosophy that the disorder in the unitary evolution of the open quantum system caused by environmental noises should be neutralized by a sequence of ordered or well-designed external operations acting on the system. This work studies the {\it exact} solution of quantum-state-diffusion equations by mixing two channels of environmental noises, i.e., relaxation (dissipation) and dephasing. It is interesting to find in two-level and three-level atomic systems, that a non-Markovian relaxation or dissipation process can be suppressed by a Markovian dephasing noise. The discovery results in an anomalous control strategy by coordinating relaxation and dephasing processes. Our approach opens up a novel revenue towards noise control strategy with no artificial manipulation over the open quantum systems.
\end{abstract}


\maketitle

\section{Introduction}

A microscopic quantum system is inevitably open due to unwanted interference from its environment~\cite{qnoise}. Even a well-isolated system has to be in risk of losing fidelity during the interaction with ancillary or interface systems~\cite{milburn}. This interaction is prerequisite to store, evolve, read, and manipulate a pure state of the interested quantum system coherently. Thus one must take exceptional care that no outside noise interferes and then the evolution of the system is precisely the desired one. Decoherence~\cite{Breuer} is the key fundamental challenge in quantum computation and quantum information processing~\cite{nielson}: to robustly store quantum states for a long time and evolve them according to specific targets of quantum engineering. Therefore, addressing the decay of an open quantum system due to external noises is desired before a practice of quantum state manipulation~\cite{miao}. Moreover, quantum noises are associated with fundamental issues in physics, such as the classical-quantum transition~\cite{Zurek}.

Various strategies have been proposed to suppress or neutralize the effects from uncontrollable quantum noises. To name a few, (i) decoherence free subspace~\cite{DFS1,DFS2,DFS3,DFS4}, which is isolated, by virtue of a dynamical symmetry, from the system-bath interaction; (ii) adiabatic elimination~\cite{ae1,ae2,ae3}, through which far-detuning levels or subspaces together with their noisy fluctuations can be removed by constructing an effective Hamiltonian resulting in an evolution of the ground states; (iii) the family of dynamical decoupling~\cite{dd1,dd2,dd3,dd4,dd5,dd6}, which dates back to the spin-echo technique, used to cancel the inhomogeneous broadening by applying $\pi$ inversion pulses. As an effective and widely employed approach, dynamical decoupling has been developed to subjectively tackle general decoherence by enforcing well-designed pulse sequences onto the evolution of open quantum systems. It is a good practice fitting well with the common philosophy -- control of disorder by order. The disordered dynamics is suppressed by ordered control, and then the decoherence is suppressed. For a more disordered dynamics of the open system induced by multiple mechanisms or channels of decoherence, more layers of dynamical decoupling control sequences might be necessary~\cite{Gong1,Gong2,Wen} in control implementation.

However, when an open system coupled to two or more sources of noises, its coherence, inner correlation or entanglement might be enhanced or reduced by the correlation between the two participating noises~\cite{BT,JRJT}. This phenomenon implies an alternative way to manipulate the system purely by noise~\cite{JW}. The physics behind this anomalous behaviour is the modified statistical property of the effective noise acting on the controlled system. This is a common point shared by noise control and dynamical decoupling, while the latter is accomplished by creating a filter function~\cite{filter1,filter2} via ordered artificial pulses. The previous investigations in the line of noise control focus on the noises of the same decoherence mechanism or those in classical systems~\cite{classical1,classical2}. In this work, we target on answering the question that to what extent the relaxation dynamics of an open quantum system could be modified by a dephasing noise and vice versa. As a further question, can one employ a dephasing noise, Markovian or non-Markovian, to interpolate the non-unitary evolution of the system embedded in a dissipative bosonic environment? Many approaches, such as master equation~\cite{Breuer} and Kraus operator decomposition~\cite{nielson}, have been exploited to understand the behavior of two-level atomic systems and three-level atomic systems in the presence of either relaxation noise or dephasing noise. Few literatures~\cite{Zhou} dealt with both dephasing and relaxation noises simultaneously, typically following the standard procedures associated with the second-order perturbation master equation~\cite{Breuer} under the Born-Markov approximation. Here we provide an exact study on open quantum systems under the presence of these two noises.

The paper is organized as follows. In Sec.~\ref{model}, we introduce the exact solution for a two-level atomic system (qubit) under the relaxation and dephasing noises simultaneously via the quantum-state-diffusion approach. In Sec.~\ref{result}, we present the numerical result of system fidelity and analysis the noise control strategy over qubit system with various combinations of noise strengths and memory capacities for both noises. It is demonstrated that the mixture of two noises along different channels can indeed be applied to defeat decoherence, especially when a nearly Markovian dephasing noise encounters a strong non-Markovian relaxation noise. Similar analysis is then extended to a $\Lambda$-type three-level atomic system to achieve a control diagram. Discussions are provided in Sec.~\ref{discuss}. We conclude this work in Sec.~\ref{conc}.

\section{Models and method}\label{model}

In this work, we apply the non-Markovian quantum-state-diffusion (QSD) equation~\cite{QSD1,QSD2,QSD3,QSD4,QSD5} to obtain an exact solution of models under both relaxation (dissipation) and dephasing channels. We first consider a microscopic model for an open qubit system. The full Hamiltonian can be written as
\begin{equation}\label{Htot}
H_{\rm tot}=\frac{\om+\xi(t)}{2}\si_z+(\si_-B^\da+h.c.)+\sum_k\om_kb_k^\da b_k,
\end{equation}
where $\om$ is the bare-frequency of the qubit system, $\xi(t)$ is a Gaussian dephasing noise satisfying $M[\xi(t)]=0$ ($M[\cdot]$ means ensemble average) and $M[\xi(t)\xi(s)]=\alpha(t-s)$ (the dephasing correlation function), $\si_z$ and $\si_\pm$ are Pauli matrices, $b_k$ and $\om_k$ are respective the annihilation operator and eigen-frequency of the $k$-th mode of the environment, $B\equiv\sum_kg_kb_k$ serves as a collective environmental operator describing the relaxation channel and $g_k$'s are the coupling strengths between system and the environmental modes.

In the rotating frame with respect to $S=\sum_k\om_kb_k^\da b_kt+\Xi(t)\frac{\si_z}{2}$, where $\Xi(t)\equiv\int_0^tds\xi(s)$, the full Hamiltonian becomes
\begin{eqnarray}\non
H_I&=&e^{iS}H_{\rm tot}e^{-iS}-\dot{S} \\ \label{HI}
&=&\frac{\om}{2}\si_z+\si_+e^{i\Xi(t)}B(t)+\si_-e^{-i\Xi(t)}B^\da(t).
\end{eqnarray}
At zero temperature, the correlation function of the relaxation environment (noise) can be represented by $\beta(t-s)\equiv\la B(t)B^\da(s)\ra=\sum_k|g_k|^2e^{-i\om_k(t-s)}$, where $B(t)\equiv\sum_kg_kb_ke^{-i\om_kt}$. In the framework of the QSD approach~\cite{QSD1,QSD2,QSD3,QSD4,QSD5}, one can obtain a stochastic Schr\"odinger equation for the system wavefunction $|\psi_t(z^*)\ra$ from the Schr\"odinger equation
\begin{equation*}
i|\dot{\Psi}(t)\ra=H_I|\Psi(t)\ra,
\end{equation*}
where $|\Psi(t)\ra$ stands for the full state of the system and the environment. Initially it is supposed that the state of the system and the environment can be factorized: $|\Psi(0)\ra=|\psi_0\ra|0\ra$. Explicitly, we define that $|\psi_t(z^*)\ra\equiv\la z||\Psi(t)\ra$, where $||z\ra=\prod_k||z_k\ra$ stands for the tensor product of stochastic Bargmann coherent states for all of the environment modes. The $||z\ra$'s constitute an over-complete set for spanning the state of the environment. Note that the reduced density matrix $\rho_{\rm sys}$ for the open system can be recovered by an ensemble averaging over these stochastic wavefunctions $|\psi_t(z^*)\ra$. After a straightforward derivation by projecting the Schr\"odinger equation to a stochastic state $\la z||$, we can achieve a formal QSD equation,
\begin{equation}\label{QSD}
\pa_t|\psi_t(z^*)\ra=[-iH_{\rm sys}+Lz_t^*-L^\da\bar{O}(t,z^*)]|\psi_t(z^*)\ra,
\end{equation}
where $H_{\rm sys}$ and $L$ denote respectively system Hamiltonian and system-environment coupling operator. Here by Eq.~(\ref{HI}), $H_{\rm sys}=\frac{\om}{2}\si_z$, $L=\si_-$. The composite and complex process $z_t^*=-i\sum_kg_k^*z_k^*e^{i\om_kt-i\Xi(t)}$ captures both (semi-classical) dephasing noise and (purely quantum mechanical) relaxation noise~\cite{qnoise}, describing the combined effect of these two noises on the nonunitary evolution of the open two-level system. These two noises are assumed to be statistically-independent. The correlation function for $z_t^*$ is then found to be
\begin{eqnarray*}
G(t-s)&=&M[z_tz_s^*] \\ &=&\sum_k|g_k|^2e^{-i\om_k(t-s)}M\left[e^{i\Xi(t)-i\Xi(s)}\right] \\ &=& \beta(t-s)e^{-\int_s^tdt_1\int_s^{t_1}dt_2\al(t_1-t_2)}.
\end{eqnarray*}
In the absence of $\xi(t)$, $G(t-s)$ is reduced to the correlation function $\beta(t-s)$ for the relaxation noise. The system operator $\bar{O}(t,z^*)$ in Eq.~(\ref{QSD}), named as the O-operator, is exactly found to be $F(t)\si_-$ by the consistency condition~\cite{QSD1,QSD2}, where the coefficient $F(t)$ is determined by $G(t-s)$:
\begin{equation}\label{cc2}
F(t)\equiv\int_0^tdsG(t-s)f(t,s),
\end{equation}
Here the coefficient function $f(t,s)$ satisfies $\pa_tf(t,s)=[i\om+F(t)]f(t,s)$ and $f(s,s)=1$. Both $F(t)$ and $f(t,s)$, as key coefficients in QSD equation~(\ref{QSD}) and following master equation~(\ref{ME}) can be numerically evaluated. After a standard derivation~\cite{QSD1,QSD2} by Eq.~(\ref{QSD}), the Novikov theorem and the definition $\rho_{\rm sys}=M[|\psi_t(z^*)\ra\la\psi_t(z^*)|]$, we can obtain an exact master equation for the central qubit system,
\begin{eqnarray}\non
\dot{\rho}_{\rm sys}&=&-i\left[\frac{\om+F_I(t)}{2}\si_z, \rho_{\rm sys}\right]\\ \label{ME} &+& F_R(t)(2\si_-\rho_{\rm sys}\si_+-\si_+\si_-\rho_{\rm sys}-\rho_{\rm sys}\si_+\si_-).
\end{eqnarray}
Here $F_R(t)$ and $F_I(t)$ are, respectively, the real and imaginary parts of $F(t)$.

The QSD equation~(\ref{QSD}) is valid for the bosonic environment with an arbitrary correlation function. For simplicity, both relaxation and dephasing noises throughout this work are described by the well-known Ornstein-Uhlenbek (OU) noise with an exponential-decay correlation function. Specifically $\beta(t-s)=\frac{\Ga_\beta\ga_\beta}{2}e^{-\ga_\beta|t-s|}$ and $\al(t-s)=\frac{\Ga_\al\ga_\al}{2}e^{-\ga_\al|t-s|}$, in which $\Ga_\beta$ and $\Ga_\alpha$ are the coupling strengths or decoherence rates and $\ga_\beta$ and $\ga_\al$ are inversely proportional to the memory capacity of the relevant noise or environment. Thus it turns out that
\begin{equation}\label{nonMarkov}
G(t-s)=\beta(t-s)\exp\left\{-\frac{\Ga_\al}{2}\left[(t-s)+\frac{e^{-\ga_\al(t-s)}-1}{\ga_\al}\right]\right\}.
\end{equation}
In general, the composite correlation function $G(t-s)$ does no longer maintain the form of linear exponential decay, i.e., two OU noises do not necessarily yield another OU noise. Unless in the Markovian limit of either noise, e.g., when $\ga_\al\rightarrow\infty$, the correlation function of the dephasing noise becomes $\al(t-s)=\Ga_\al\de(t-s)$, Eq.~(\ref{nonMarkov}) then reduces to
\begin{equation}\label{Markov}
G(t-s)=\frac{\ti{\Ga}_\beta{\ti\ga}_\beta}{2}e^{-{\ti\ga}_\beta|t-s|},
\end{equation}
where $\ti{\ga}_\beta=\ga_\beta+\Ga_\al/2$, $\ti{\Ga}_\beta=r\Ga_\beta$ and $r=\ga_\beta/\ti{\ga}_\beta$. Due to the fact that $\ti{\ga}_\beta>\ga_\beta$ and $\ti{\Ga}_\beta<\Ga_\beta$, we now have a modified OU noise described by Eq.~(\ref{Markov}). Comparing to the original relaxation noise described by $\beta(t-s)$, the effective memory time (correlation time) for the composite noise becomes shorter, while the strength becomes smaller. In the short-time limit, the decoherence rate of the qubit system is cut down. But in a moderate time-scale, the probability of restoring the qubit to its original state by the back-action of the environment is smaller than before. The competition between these two effects would give rise to interesting behaviors in the control dynamics.

Using the exact equation of motion~(\ref{QSD}) or (\ref{ME}) and the initial state of the two-level atomic system $|\psi_0\ra=\mu|1\ra+\nu|0\ra$, where $|\mu|^2+|\nu|^2=1$, the fidelity $\mathcal{F}(t)\equiv\la\psi_0|\rho_{\rm sys}(t)|\psi_0\ra$ of the open system during the nonunitary evolution is formally found to be
\begin{eqnarray}\non
\mathcal{F}(t)&=&1-|\mu|^2-(|\mu|^2-2|\mu|^4)e^{-2\int_0^tds{\rm Re}[F(s)]} \\ \label{qubitF} &+& 2(|\mu|^2-|\mu|^4){\rm Re}\left[e^{-\int_0^tdsF(s)}\right],
\end{eqnarray}
where ${\rm Re}[X]$ means the real part of $X$. It is found that the fidelity depends on the initial population of the system and the coefficient $F(t)$ in Eqs.~(\ref{QSD}) or (\ref{ME}). Due to Eqs.~(\ref{cc2}) and (\ref{Markov}), we have
\begin{equation*}
\dot{F}(t)=\frac{\ti{\Ga}_\beta{\ti\ga}_\beta}{2}+(-{\ti\ga}_\beta+i\om)F(t)+F^2(t).
\end{equation*}
Note that $F(0)=0$ according to Eq.~(\ref{cc2}).

\section{Results of fidelity control}\label{result}

\subsection{Two-level atomic system}\label{twolevel}

To demonstrate the effect of the noise mixing in a broader perspective, we calculate the average fidelity $\bar{\mathcal{F}}$ over all pure initial states of the qubit system by taking $\mu=\cos\frac{\theta}{2}$ and $\nu=\sin\frac{\theta}{2}e^{i\phi}$ and integrating over $\theta\in[0, \pi]$ and $\phi\in[0, 2\pi]$. From Eq.~(\ref{qubitF}), we have
\begin{eqnarray}
\bar{\mathcal{F}}(t)&=&\frac{1}{4\pi^2}\int_0^{\pi}\sin\theta d\theta\int_0^{2\pi}d\phi \mathcal{F}(t)=\frac{1}{2} \\ \non &+&\frac{\exp\left(-2\int_0^tds{\rm Re}[F(s)]\right)+{\rm Re}\left[\exp\left(-\int_0^tdsF(s)\right)\right]}{4}.
\end{eqnarray}

\begin{figure}[htbp]
\centering
\subfigure{\label{twoMar:gammab0.1}
\includegraphics[width=2.8in]{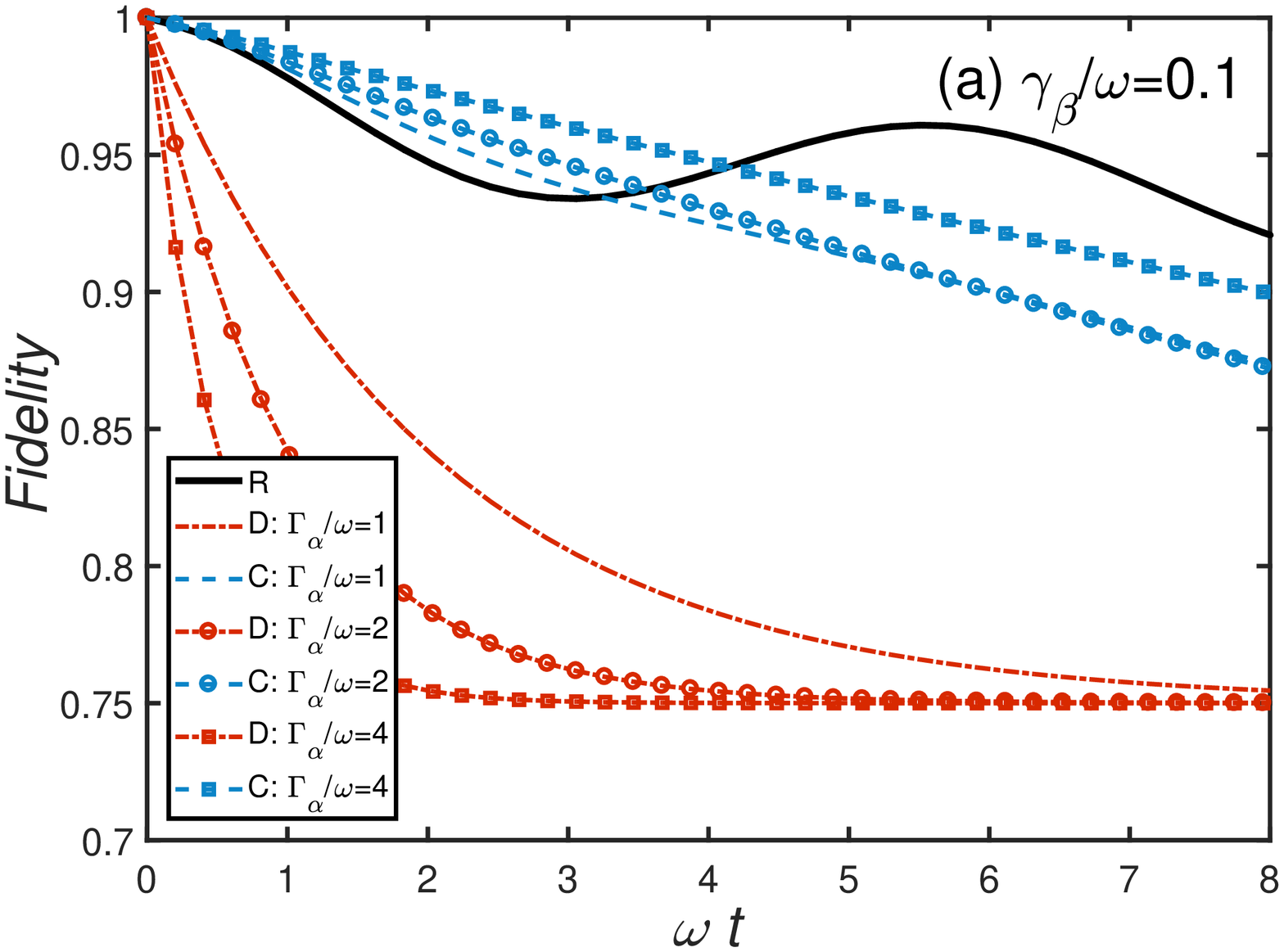}}
\subfigure{\label{twoMar:gammab0.5}
\includegraphics[width=2.8in]{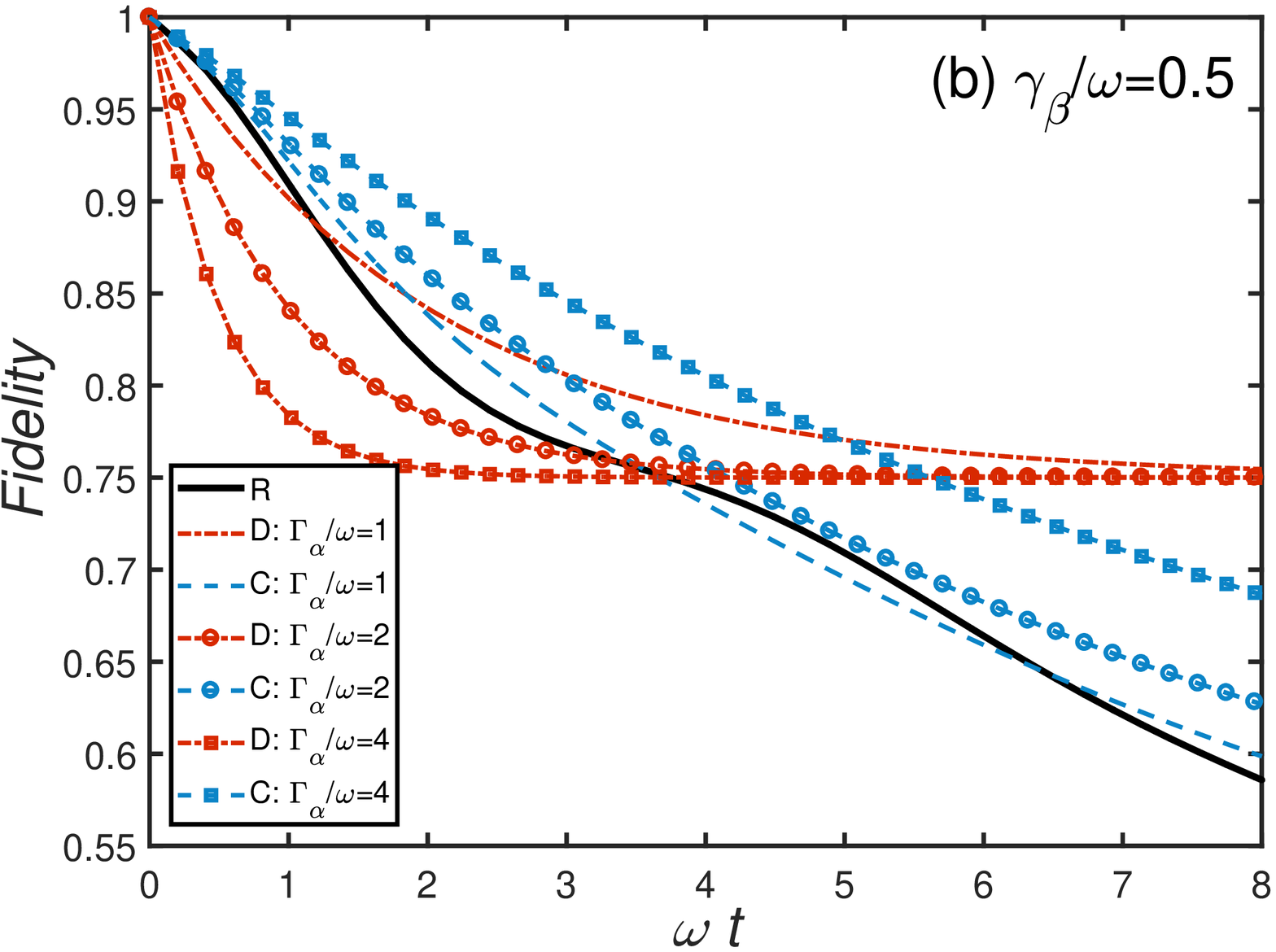}}
\subfigure{\label{twoMar:gammab2.0}
\includegraphics[width=2.8in]{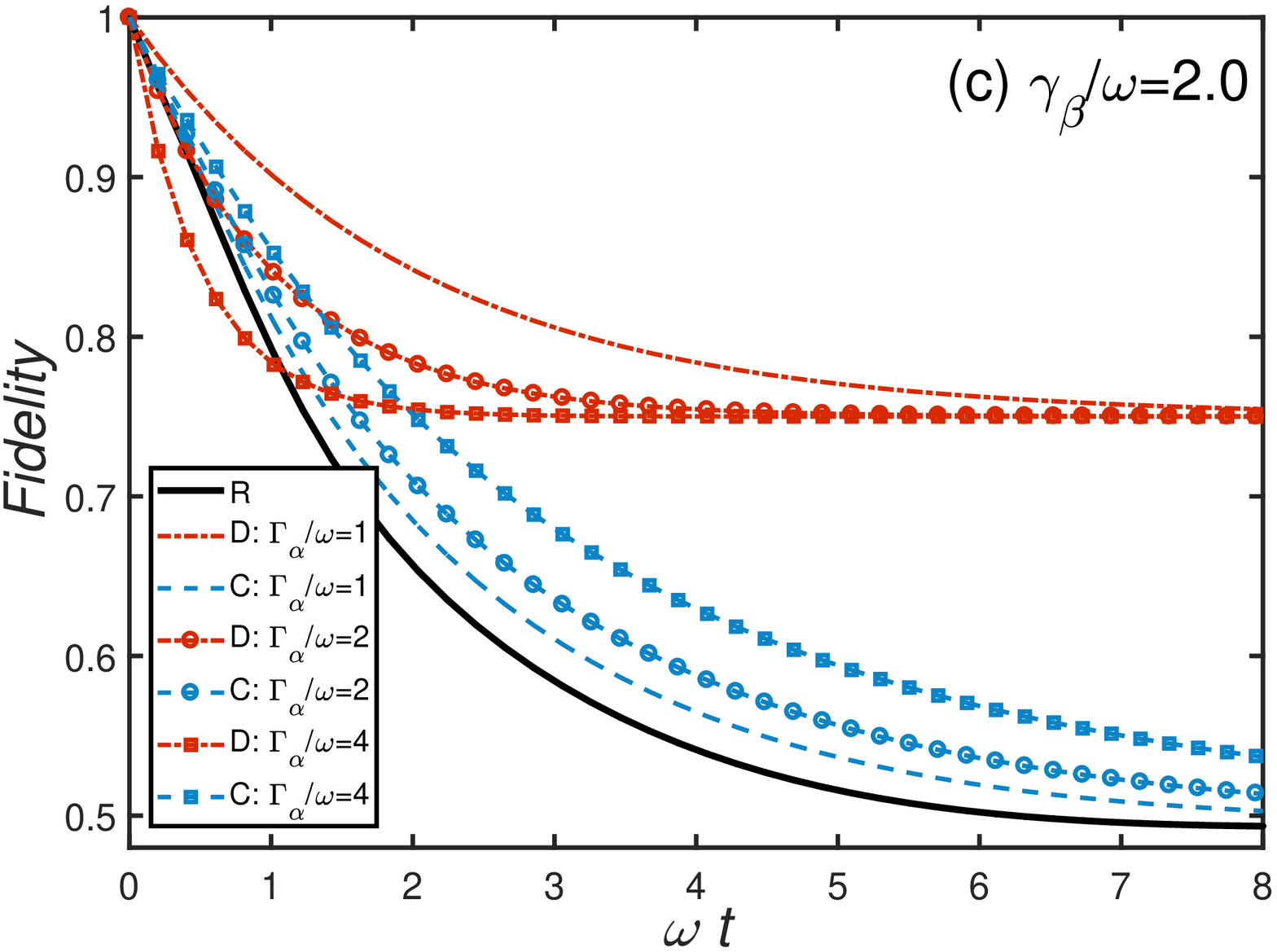}}
\caption{(Color online) Average fidelity $\bar{\mathcal{F}}$ of a two-level atomic system in the presence of a general relaxation noise and a Markovian dephasing noise under different parameters. $R$, $D$ and $C$ represent the dynamics under pure relaxation noise, pure dephasing noise and a mixture of noises, respectively. We choose $\Gamma_\beta/\omega=1$. (a) $\gamma_\beta/\omega=0.1$; (b) $\gamma_\beta/\omega=0.5$; (c) $\gamma_\beta/\omega=2.0$.}\label{twoMar}
\end{figure}

We first consider the effect of Markovian dephasing noise on the relaxation process with different memory capacity parameterized by $\ga_\beta$. For each subfigure in Fig.~\ref{twoMar}, we fix the noise strength of the relaxation process $\Ga_\beta$ and enforce the dephasing noise with different noise strength $\Ga_\al$ onto the system. The black-solid curve labelled by $R$ represents the average fidelity dynamics of the qubit system experiencing a purely relaxation noise. The red-dot-dashed curves labelled by $D$ represent the fidelity dynamics driven by a purely dephasing noise, while the blue-dashed curves labelled by $C$ demonstrate the fidelity under the mixed noises. Along the red and blue curves, different marks are used to denote different strengths (decoherence rates) of the dephasing noise and respective component in noise mixture.

In Fig.~\ref{twoMar}, the three subfigures are arranged by the increasing $\ga_\beta$, i.e., by the decreasing memory capacity of the relaxation noise. With a strongly non-Markovian relaxation noise implied by $\ga_\beta/\om=0.1$, Fig.~\ref{twoMar:gammab0.1} can be clearly divided into two regions for each triple of curves grouped by the black curve and the red and blue curves indicating the same value of pure dephasing rate $\Ga_\al$. In a moderate time scale $0<\om t<\tau$, with a critical value $\tau$ around $3$ to $4$, the average fidelity under the mixture of noises is higher than that purely under either relaxation or dephasing noise. A larger dephasing noise strength $\Ga_\al$ leads to a longer critical time $\tau$ and a higher average fidelity. While in a longer time scale after the critical moment $\tau$, the fidelity under the mixture of noises $C$ is still higher than that under pure dephasing $D$; however, it is lower than the fidelity under pure relaxation $R$. It means that our noise-control protocol works in the short time limit.

With a moderate non-Markovian relaxation noise implied by $\ga_\beta/\om=0.5$, the whole pattern in Fig.~\ref{twoMar:gammab0.5} becomes involved. In the presence of a comparatively weak dephasing strength $\Ga_\al/\om=1$, it is hard to distinguish these three curves (the black-solid curve and the red-dot-dashed and blue-dashed curves without marks) in a short time scale $0<\om t<2$. Over that interval, the decoherence rate under the mixture of noises is larger than that under the dephasing noise and still almost the same as that under the pure relaxation noise. When we have a moderate dephasing strength $\Ga_\al/\om=2$, the fidelity is enhanced from the standpoint of the pure relaxation process and maintains an advantage over that under the dephasing noise in $0<\om t<4$. When the dephasing rate is increased to $\Ga_\al/\om=4$, the time scale in which the noise-control strategy works will be extended to about $0<\om t<5.5$.

Figure~\ref{twoMar:gammab2.0} is used to demonstrate the mixture of a nearly Markovian relaxation noise $\ga_\beta/\om=2.0$ and a Markvoian dephasing noise. It is shown that the effective decay rate based on the pure relaxation noise can be always mitigated by the ``addition'' of a dephasing noise. While only with a quite strong dephasing noise $\Ga_\alpha/\om=4.0$, the fidelity under mixed noises is higher than that under the pure-dephasing noise in a very limited time scale $0<\om t<1.8$.

Therefore, by using the Markovian dephasing noise as a control tool, the average fidelity of the relaxation process can be generally enhanced during a long time evolution. However, the value of the fidelity should be over a threshold for any practical quantum information processing. That can be only realized in the strong non-Markovian limit. In Fig.~\ref{twoMar:gammab0.1}, the fidelity is improved to be over $0.95$ in the time scale $0<\om t<4$ by a Markovian dephasing noise with a large strength $\Ga_\alpha/\om=4$ (see the blue-dashed curve with square marks). From the viewpoint of an open quantum system under a Markovian dephasing process, its average fidelity can also be increased via the ``addition'' of a relaxation noise. The effect of noise-control scheme becomes more significant in the presence of a relaxation noise with a longer memory time.

\begin{figure}[htbp]
\centering
\subfigure{\label{twonM:gammab}
\includegraphics[width=2.8in]{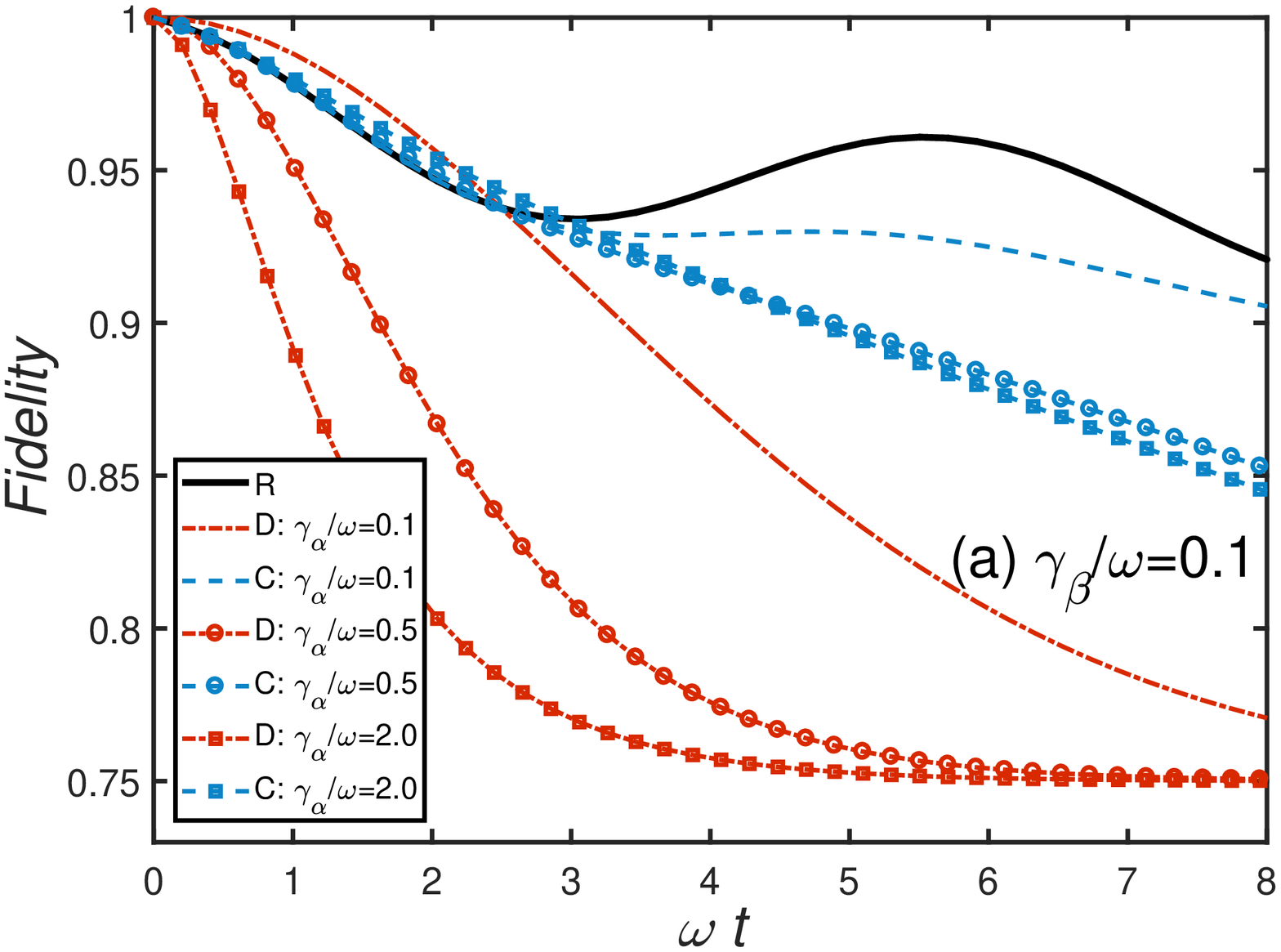}}
\subfigure{\label{twonM:gammaa}
\includegraphics[width=2.8in]{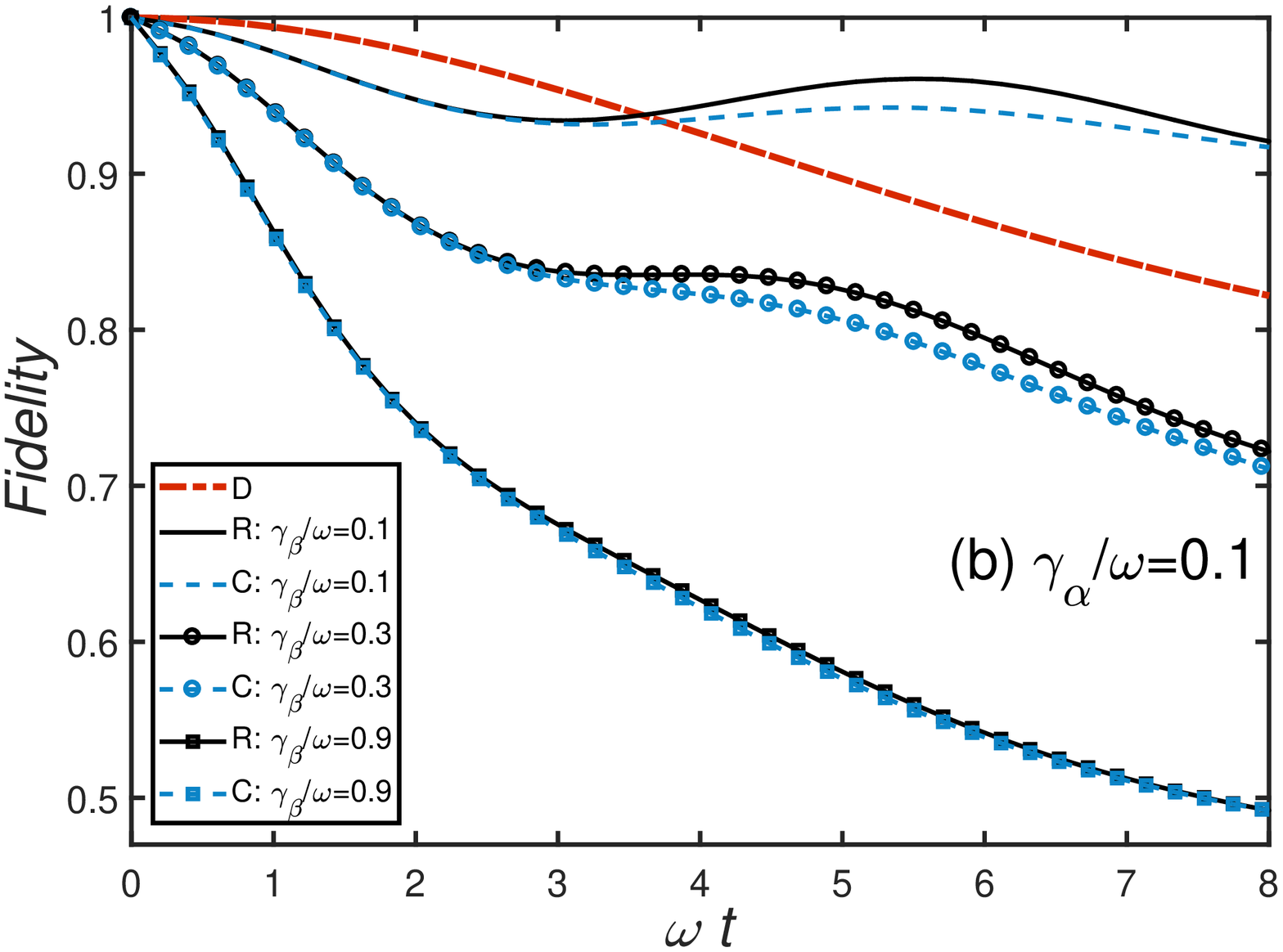}}
\caption{(Color online) Average fidelity $\bar{\mathcal{F}}$ of a two-level atomic system in the presence of a non-Markovian relaxation noise and a non-Markovian dephasing noise under different parameters. $R$, $D$ and $C$ represent the dynamics under pure relaxation noise, pure dephasing noise and mixture of noise, respectively. We choose $\Gamma_\beta/\omega=1$. (a) $\gamma_\beta/\omega=0.1$ and $\Gamma_\alpha/\omega=2.0$; (b) $\gamma_\alpha/\omega=0.1$ and $\Gamma_\alpha/\omega=1.0$.}\label{twonM}
\end{figure}

To see the effect of mixing two non-Markovian noises, we come back to a more general situation described by the composite correlation function given in Eq.~(\ref{nonMarkov}). The numerical results are presented in Fig.~\ref{twonM} to compare the evolutions of an open qubit system under two OU noises indicated respectively by $\Ga_\alpha$, $\ga_\alpha$, and $\Ga_\beta$, $\ga_\beta$ and that under a non-OU noise with a square of time-difference in the exponential correlation function.

In Fig.~\ref{twonM:gammab}, it turns out that the addition of a non-Markovian dephasing noise with various memory parameters plays almost no positive role in enhancing the average fidelity of the qubit state in the presence of a fixed strong non-Markovian relaxation noise with $\ga_\beta/\om=0.1$. The fidelity under the mixture of the two noises is nearly larger than that under the pure dephasing noise. The application of the dephasing noise-control does not significantly change the relaxation rate in a short time scale $0<\om t<3$, but remarkably accelerates the decoherence rate in a long time scale $\om t>3$.

Then in Fig.~\ref{twonM:gammaa}, the participating dephasing noise is fixed with a long memory time implied by $\ga_\alpha/\om=0.1$. It is demonstrated that the presence of the dephasing noise contributes little to the evolution determined by the relaxation noise. When $\ga_\beta/\om=\ga_\alpha/\om=0.1$, the fidelity under the mixture of two noises is higher than that under the pure dephasing noise in the time scale $0<\om t<3.6$. When $\ga_\beta/\om$ is increased to $0.3$, i.e., the memory capacity of the relaxation noise is weaker than that of the dephasing noise, the fidelity under the mixture of two noises is lower than those purely under either of them. Moreover, when $\ga_\beta/\om=0.9$, the fidelity under the mixture of two noises is almost the same as that under the pure relaxation noise, i.e., the effect of the additional dephasing noise disappears.

The results in Figs.~\ref{twoMar} and \ref{twonM} strongly indicate a control principle: to mutually cancel two unwanted noisy processes enforced into the unitary evolution of the quantum system, the characteristic time scales of these two processes should be scale-separable. Particularly in our model, the characteristic time scales of the two mixed noises could be inferred by the memory parameters $\ga_\al$ and $\ga_\beta$, respectively. When either of them approaches zero, the corresponding noise signals have a long-time-scale correlation, which means its characteristic time scale is quite short and the noise evolves in a rapid frequency in the time domain. When either of them approaches infinity, the corresponding noise signals are uncorrelated in the time domain, which means its characteristic time scale also approaches infinity and its evolution is slow. Upon the quantum Zeno effect~\cite{Zeno1,Zeno2,Zeno3}, the survival probability of the initial state (state fidelity) approaches unity when the measurement period is far less than the Zeno time. The fast-frequent measurement sequence can then maintain the state fidelity by defeating the slowly-varying decoherence noise. Yet here it is worth emphasising that our noise-control protocol involves no quantum measurement, so it is different from the quantum Zeno effect despite sharing a similar control principle with it. Our protocol stems from the effective correlation function for the mixed noises instead of regular or random sequence of measurements. Another interesting point in this work is that, our protocol performs well with a strong non-Markovian relaxation noise and a Markovian dephasing noise, but it does not work with a Markovian relaxation noise and a non-Markovian dephasing noise. The latter situation can be referred to by the blue-dashed curve with the square marks in Fig.~\ref{twonM:gammaa}.

\subsection{Three-level atomic system}

We find that the result and principle of the above control are also applicable to systems with higher degrees of freedom. Yet for more complicated systems, it is hard to obtain an exact analytical expression of fidelity for arbitrary initial states of the open system via a quantum state diffusion study. To further confirm the validity of our noise-control protocol, we provide a control diagram in this subsection. We now consider a $\Lambda$-type three-level atom~\cite{QSD5,Leakage} with system Hamiltonian $H_{\rm sys}=\frac{\om}{2}(|1\ra\la1|-|2\ra\la2|-|3\ra\la3|)$ and system-environment coupling operator $L_{\Lambda}=|2\ra\la1|+|3\ra\la1|$. Considering the results in the case of the two-level system, we investigate the mixture of a non-Markovian relaxation noise and a Markovian dephasing noise as described by the composite correlation function in Eq.~(\ref{Markov}), since this choice could demonstrate the advantage of composite noises over single-component noise in the most efficient way. The total Hamiltonian shares the same form as that in Eq.~(\ref{Htot}) by replacing $\si_z$ and $\si_-$ with $|1\ra\la1|-|2\ra\la2|-|3\ra\la3|$ and $L_{\Lambda}$, respectively. As for the formal QSD equation~(\ref{QSD}), the operator $L$ is now written as $L_{\Lambda}$ and the exact O-operator in this model is found to be $\bar{O}(t,z^*)=Q(t)L_{\Lambda}$ with a coefficient function $Q(t)$. After a similar derivation from Eq.~(\ref{QSD}) to Eq.~(\ref{ME}) for qubit system, we can obtain an exact master equation for this $\Lambda$-type three-level atomic system in the rotating frame with respect to $H_{\rm sys}$:
\begin{equation*}
\dot{\rho}_{\rm sys}=Q(t)[L_{\Lambda}\rho_{\rm sys}, L^\da_{\Lambda}]+Q^*(t)[L_{\Lambda}, \rho_{\rm sys}L^\da_{\Lambda}].
\end{equation*}
Consequently, assuming the atom is initially prepared as a normalized state $|\psi_0\ra=a|1\ra+b|2\ra+c|3\ra$, we have
\begin{eqnarray}\non
\mathcal{F}_{\Lambda}(t)&=&\frac{1}{2}|a|^2[1-e^{\bar{Q}(t)+\bar{Q}^*(t)}]
|b+c|^2+|a|^4e^{\bar{Q}(t)+\bar{Q}^*(t)}\\ \label{fidelity3l} &+&(1-|a|^2)^2+(|a|^2-|a|^4)\left[e^{\bar{Q}(t)}+e^{\bar{Q}^*(t)}\right],
\end{eqnarray}
where $\bar{Q}(t)\equiv-2\int_0^tdsQ(s)$ and $Q(t)$ satisfies
\begin{equation*}
\dot{Q}(t)=\frac{\ti{\Ga}_\beta{\ti\ga}_\beta}{2}+(-{\ti\ga}_\beta+i\om)Q(t)+2Q^2(t),
\end{equation*}
with $Q(0)=0$.

\begin{figure}[htbp]
\centering
\includegraphics[width=3in]{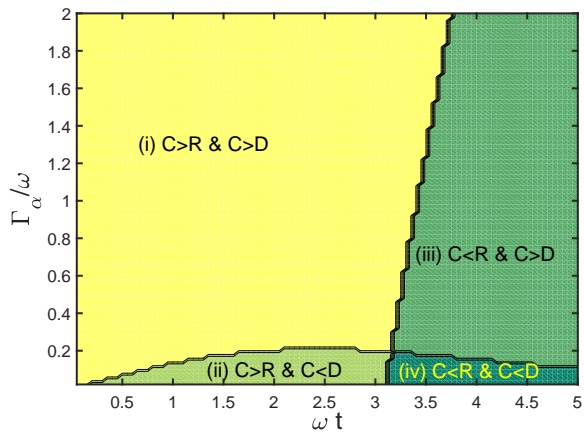}
\caption{(Color online) Comparison of fidelities for a $\Lambda$-type three-level atomic system in the space of evolution time-dephasing rate: $\omega t-\Gamma_\alpha/\omega$. The fidelity $\mathcal{F}_{\Lambda}(t)$ is obtained by Eq.~(\ref{fidelity3l}) in the presence of a non-Markovian relaxation noise and a Markovian dephasing noise. We choose the initial state as $|\psi_0\ra=(|1\ra+|2\ra)/\sqrt{2}$. The parameters for relaxation noise are $\Gamma_\beta/\omega=1$ and $\gamma_\beta/\omega=0.1$. This diagram is divided into four regions according to the relations among the fidelities under composite noise $C$, of single relaxation noise $R$, and of single dephasing noise $D$, respectively. } \label{threelevel}
\end{figure}

We plot a diagram in Fig.~\ref{threelevel} to demonstrate the relations among the fidelities under composite and individual noises, where the initial state of this system is chosen as $|\psi_0\ra=(|1\ra+|2\ra)/\sqrt{2}$. As in subsection~\ref{twolevel}, $R$, $D$ or $C$ labels the fidelities of the open system under respectively relaxation noise, dephasing noise, and the mixture of these two noises.

The whole parameter space is divided into four regions with different colors, by which one can capture a complete picture of the control effect in quality. The yellow region (i) indicates the parameter subspace in which the fidelity in the presence of the mixture noise is higher than those in the presence of either individual noises. We label it by $C>R$ and $C>D$. It shows that the working width of the time scale windows for our noise-control strategy increases steadily with the strength of the dephasing noise $\Ga_{\alpha}$. The two green regions (iii) and (iv) imply two failure conditions for the noise-mixing strategy. In the long time scale, the fidelity under composite noise is lower than that under the dissipation noise. Similar to Fig.~\ref{twonM:gammaa}, there is even a small region (iv) in the parameter space where the fidelity under composite noise is lower than those under either individual noises. More than Fig.~\ref{twonM:gammaa}, in this case study of the three-level atom, we find our control strategy would surely fail in the long time scale when the strength of the Markovian dephasing noise $\Ga_\al$ is too weak. Before this happens, we observe a small region (ii) in which the fidelity under composite noise is lower than that under the dephasing noise, but higher than that under the dissipation noise. Therefore, our noise-control protocol for the three-level system is as reliable as the application in Fig.~\ref{twoMar} for the two-level system, that a strong Markovian dephasing noise can be used to alleviate the decoherence caused by a non-Markovian dissipation noise for quantum systems with a higher dimensionality.

\section{Discussion}\label{discuss}

The full Hamiltonian we considered in Eq.~(\ref{Htot}) can be derived by a pure quantum model with dephasing and dissipation due to the same environment:
\begin{equation}\label{Htot2}
H_{\rm tot}=\frac{\om}{2}\si_z+\si_z(B^\da+B)+(\si_-B^\da+h.c.)+\sum_k\om_kb_k^\da b_k,
\end{equation}
or due to two separable environments:
\begin{eqnarray}\non
H_{\rm tot}&=&\frac{\om}{2}\si_z+\si_z(B^\da_1+B_1)+\sum_k\om_{k1}b_{k1}^\da b_{k1} \\ \label{Htot3} &+&(\si_-B^\da_2+h.c.)+\sum_k\om_{k2}b_{k2}^\da b_{k2},
\end{eqnarray}
where $B_{n}\equiv\sum_kg_{kn}b_{kn}$. When $B^\da+B$ or $B^\da_1+B_1$ is treated as a classical stochastic variable $\xi(t)/2$, Eq.~(\ref{Htot2}) or (\ref{Htot3}) is reduced to Eq.~(\ref{Htot}). In this way, we can study simultaneously a classical dephasing noise and a quantum relaxation noise. Either classical or quantum noise model has been well considered in previous literatures, but here we study a combination of them.

It is well-accepted that the environmental correlation function as well as spectral function play important roles in determining the nonunitary evolution of the open quantum system. We employ the Ornstein-Uhlenbek form for both dephasing and dissipation noises to clearly illustrate the resulting effective environmental correlation function~(\ref{nonMarkov}). More importantly, it is convenient to demonstrate that a high-frequent noise (Markovian dephasing noise) can be used to suppress the slowly-varying decoherence process induced by a non-Markovian dissipation noise [See Figs.~\ref{twoMar:gammab0.1} and \ref{twoMar:gammab0.5}]. Our protocol can be extended to other forms of noise case by case. Many noises, such as $1/f$ noise~\cite{oneoverf} and those with spectral density in the form of a power law up to a cut-off frequency~\cite{Strunz}, can be mimicked by a summation of OU noises.

Our model can be realized physically by atomic gas represented by simple two-level or three-level systems. Spontaneous emission gives rise to the relaxation process while random recoils of the atoms cause the dephasing process. Our study reveals that a prohibitive relaxation could be induced by proper heating of the gas, which improves the random collision probability among atoms but does not significantly modify the spontaneous emission. This is not the unique way to reach the mutual-cancellation of the noises along two quantum channels. Another realization of an open quantum system model is an electron spin in a quantum dot (see e.g., Ref.~\cite{QD} and references therein), in which the central electron spin is relaxed by the surrounding phonons and flip-flop interaction with the nuclear spins. The dephasing process is easily generated by the fluctuation along the Zeeman splitting determined by external magnetic field.

\section{Conclusion}\label{conc}

In conclusion, it is well accepted that the interaction between an atom and the interference or ancillary system, e.g., the electromagnetic field outside, plays a crucial role in developing our understanding of light-matter interaction, and is central to various quantum technologies, including lasers and many quantum computing architectures. While unwanted system-environment interaction or uncontrollable noise is believed to be a negative element that is responsible for ruining coherence, accelerating the decoherence rate and causing leakage of the required encoded information. However, upon exact evaluation of the effect of coexistence of relaxation and dephasing noises, we have observed unconventional mutual cancellation of noises in the dynamics of open quantum systems. Our results provide a new direction of research on decoherence suppression in quantum control by virtue of saving source to generate well-designed control sequence. Furthermore it implies that multiple decoherence channels do not necessarily push an open quantum system even closer to the boundary of quantum-classical transition.

\section*{Acknowledgments}
We acknowledge grant support from the Basque Government (Grant No. IT986-16), the Spanish MICINN (Grant No. FIS2015-67161-P), the National Science Foundation of China (Grant No. 11575071 and 11774022), the NSAF (Grant No. U1530401), and HK PolyU (Grant No. 1-ZVGH).

\end{document}